\documentclass[12pt]{article}
 
\def\beq{\begin{equation}}
\def\eeq{\end{equation}}
\def\beqa{\begin{eqnarray}}
\def\eeqa{\end{eqnarray}}

\newlength{\dinwidth} \newlength{\dinmargin}
\begin{document}
 
\begin{center}
{\Large \bf Soft-gluon expansions through NNNLO}
\end{center}
\vspace{2mm}
\begin{center}
{\large Nikolaos Kidonakis}\\
\vspace{2mm}
{\it Kennesaw State University, Physics  \#1202\\
1000 Chastain Rd., Kennesaw, GA 30144-5591, USA}\\
\end{center}
 
\begin{abstract}
I present universal master formulas for soft-gluon corrections to
hard-scattering cross sections 
through next-to-next-to-next-to-leading order (NNNLO).
I also briefly discuss applications to some processes where
these corrections enhance the cross section and decrease the scale dependence.
\end{abstract}

\section{Soft-gluon resummation}

Cross sections in perturbative QCD can be calculated by employing
factorization theorems as 
$\sigma=\sum_f \int [ \prod_i  dx_i \, \phi_{f/h_i}(x_i,\mu_F)]\,
{\hat \sigma}(s,t,u,\mu_F,\mu_R)$ ,
where $\hat \sigma$ is the perturbatively calculable
hard-scattering cross section, and the parton distributions
$\phi$ are determined from experiment.
The renormalization and factorization scales are denoted by 
$\mu_R$ and $\mu_F$ respectively, and $s,t,u$ are standard
kinematical invariants formed from the momenta of the partons
in the hard scattering.
                                                                           
Near threshold for the production of a specified system,
such as a top quark pair or a Higgs boson, there
is restricted phase space for real gluon emission.
The incomplete cancellation of infrared divergences
between real and virtual graphs results in the
appearance of large logarithms.
If we  define $s_4=s+t+u-\sum m^2$, with $m$ the masses of the
particles in the scattering,  then $s_4 \rightarrow 0$ at threshold
and these soft-gluon logarithmic 
corrections take the form of plus distributions, 
${\cal D}_l(s_4)\equiv[\ln^l(s_4/M^2)/s_4]_+$,
where $M$ is a relevant hard scale, such as the mass of a heavy quark 
or the transverse momentum of a jet, and 
$l \le 2n-1$ for the $n$-th order corrections.

If we define moments of the cross section 
$\hat{\sigma}(N)=\int_0^{\infty} ds_4 \,
e^{-Ns_4/M^2} \; {\hat\sigma}(s_4)$
then the soft corrections are transformed as 
$[\ln^{l}(s_4/M^2)/s_4]_+$
$\rightarrow [(-1)^{l+1}/(l+1)]\ln^{l+1}N +\cdots$.
We can formally resum these logarithms $\ln N$ to all orders in $\alpha_s$ by
factorizing the  soft gluons from the hard scattering \cite{KS,LOS}.
Although the formal resummation in moment space is well defined, when 
inverting back to momentum space we encounter ambiguities 
due to the infrared singularity which require a prescription.
Unfortunately different prescriptions can give different numerical results 
as well as have dubious theoretical underpinnings 
(see discussion in Ref. \cite{NKtop}).
However, fixed-order expansions can provide us with solid,
prescription-independent, theoretical and numerical results 
\cite{NKtop,NKuni}. 
 
At next-to-leading order (NLO) in $\alpha_s$, 
the cross section includes ${\cal D}_1(s_4)$ terms which are
the leading logarithms (LL), and ${\cal D}_0(s_4)$ terms which 
are the next-to-leading logarithms (NLL).  
At next-to-next-to-leading order (NNLO), we 
have ${\cal D}_3(s_4)$ (LL), ${\cal D}_2(s_4)$ (NLL), 
${\cal D}_1(s_4)$ (NNLL), and ${\cal D}_0(s_4)$ (NNNLL) terms. 
At next-to-next-to-next-to-leading order (NNNLO), 
we have ${\cal D}_5(s_4)$ (LL), ${\cal D}_4(s_4)$ (NLL), 
${\cal D}_3(s_4)$ (NNLL), ${\cal D}_2(s_4)$ (NNNLL), 
${\cal D}_1(s_4)$ (NNNNLL), and ${\cal D}_0(s_4)$ (NNNNNLL) terms.

The threshold resummation formalism has been applied by now to many 
processes in hadron-hadron and lepton-hadron colliders, 
for both total and differential cross sections, in both 
single-particle-inclusive (1PI)  and pair-invariant-mass (PIM) kinematics, 
for both simple and complex color flows,
and in both  ${\overline{\rm MS}}$ and DIS factorization 
schemes \cite{NKuni}.

Specific processes for which soft-gluon corrections have been calculated
at NNLO include
top quark pair hadroproduction \cite{NKtop,NKttb},
beauty and charm production \cite{NKbc}, jet production \cite{NKjet},
direct photon production \cite{NKph},
large-$p_T$ $W$ production \cite{NKw},
FCNC top production \cite{NKfcnc},
and charged Higgs production \cite{NKch}.
Numerical results show that usually the 
soft corrections are a good approximation of the full NLO result.
In all cases the higher-order corrections are sizable and produce a
dramatic decrease of the scale dependence of the cross section.

The resummed cross section can be written for an arbitrary process as
\cite{NKuni}
\beqa
{\hat{\sigma}}^{res}(N) &=&
\exp\left[ \sum_i E_i(N_i)\right] \;
\exp\left[ \sum_j E'_j(N_j)\right] \;
\exp\left[2\, d_{\alpha_s} \int_{\mu_R}^{\sqrt{s}}\frac{d\mu'}{\mu'}
\beta\left(\mu'\right)\right]
\nonumber\\ && \hspace{-20mm} \times \,
\exp \left[\sum_i 2\int_{\mu_F}^{\sqrt{s}} {d\mu' \over \mu'}
\left(\frac{\alpha_s(\mu')}{\pi}\gamma_i^{(1)}
+{\gamma'}_{i/i}\left(\mu'\right)\right)\right]
\nonumber\\ && \hspace{-30mm} \times
{\rm Tr} \left \{H\left(\mu_R\right) \;
\exp \left[\int_{\sqrt{s}}^{{\sqrt{s}}/{\tilde N_j}}
{d\mu' \over \mu'} \Gamma_S^\dagger\left(\mu'\right)\right] \;
S \left(s/{\tilde N_j}^2 \right) \;
\exp \left[\int_{\sqrt{s}}^{{\sqrt{s}}/{\tilde N_j}}
{d\mu' \over \mu'}\; \Gamma_S
\left(\mu'\right)\right] \right\}
\label{resHS}
\eeqa
where $E_i(N_i)$ denotes contributions from the incoming partons
and is given in the $\overline{\rm MS}$ scheme by
\beq
E_i(N_i)=
-\int^1_0 dz \frac{z^{N_i-1}-1}{1-z}\;
\left \{\int^{\mu_F^2}_{(1-z)^2s} \frac{d\mu'^2}{\mu'^2}
A_i(\mu')
+{\nu}_i\left[(1-z)^2 s\right]\right\}
\eeq
with $A_i = C_i \, [ {\alpha_s/\pi}+({\alpha_s/\pi})^2 K/2]
+\cdots$,
${\nu}_i=(\alpha_s/\pi)C_i+(\alpha_s/\pi)^2 {\nu}_i^{(2)}
+\cdots$;
$E_j(N_j)$ denotes contributions from massless final-state partons (if any)
and is given by 
\beq
E'_j(N_j)=
\int^1_0 dz \frac{z^{N_j-1}-1}{1-z}\;
\left \{\int^{1-z}_{(1-z)^2} \frac{d\lambda}{\lambda}
A_j\left(\lambda s\right)
-B'_j\left[(1-z)s\right]
-{\nu}_j\left[(1-z)^2 s\right]\right\}
\eeq
where $B'_j=(\alpha_s/\pi){B'}_j^{(1)}+(\alpha_s/\pi)^2 {B'}_j^{(2)}+\cdots$;
$\gamma_i$ are parton anomalous dimensions;
$H$ are hard scattering matrices, independent of soft-gluon radiation;
$S$ are soft matrices which describe noncollinear soft-gluon emission;
and $\Gamma_S$ are soft anomalous dimension matrices which appear in the
evolution of the $S$ matrices. $H$, $S$, and $\Gamma_S$ are matrices 
in the space of color exchanges; they become simple functions for processes
with simple color structure, such as Drell-Yan production. 

Expansions of the resummed cross section through NNLO were given
in Ref. \cite{NKuni}, where master formulas were derived and then used
in calculations for specific processes [5-11].
Here we extend this expansion to NNNLO.

\section{NNNLO soft-gluon expansions and applications}

Expanding Eq. (\ref{resHS}) to NLO, gives us the master formula
for the NLO corrections
\beqa
{\hat{\sigma}}^{(1)} &=& \sigma^B \frac{\alpha_s(\mu_R^2)}{\pi}
\left\{c_3\, {\cal D}_1(s_4) + c_2\,  {\cal D}_0(s_4)
+c_1\,  \delta(s_4)\right\}
\nonumber \\ &&
{}+\frac{\alpha_s^{d_{\alpha_s}+1}(\mu_R^2)}{\pi}
\left[A^c \, {\cal D}_0(s_4)+T_1^c \, \delta(s_4)\right]
\eeqa
with
$c_3=\sum_i 2 \, C_i -\sum_j C_j$, where 
for quarks $C_F=(N_c^2-1)/(2N_c)$ and 
for gluons $C_A=N_c$;
$c_2=c_2^{\mu}+T_2$,
with
$c_2^{\mu}=-\sum_i C_i \ln(\mu_F^2/M^2)$
and
\beq
T_2=- \sum_i \left[C_i
+2\,  C_i \,  \ln\left(\frac{-t_i}{M^2}\right)+
C_i \, \ln\left(\frac{M^2}{s}\right)\right]
-\sum_j \left[{B'}_j^{(1)}+C_j
+C_j \, \ln\left(\frac{M^2}{s}\right)\right] \, ;
\eeq
and
$c_1 =c_1^{\mu} +T_1$, with
\beq
c_1^{\mu}=\sum_i \left[C_i\, \ln\left(\frac{-t_i}{M^2}\right)
-\gamma_i^{(1)}\right]\ln\left(\frac{\mu_F^2}{M^2}\right)
+d_{\alpha_s} \frac{\beta_0}{4} \ln\left(\frac{\mu_R^2}{M^2}\right) \, ,
\eeq
where  
for quarks ${B'}_q^{(1)}=\gamma_q^{(1)}=3 C_F/4$
and
for gluons ${B'}_g^{(1)}=\gamma_g^{(1)}=\beta_0/4$.
The term $A^c$  involves matrices and is given by
$A^c={\rm tr} \left(H^{(0)} {\Gamma'}_S^{(1)\,\dagger} S^{(0)}
+H^{(0)} S^{(0)} {\Gamma'}_S^{(1)}\right)$.
Finally $T_1$ and $T_1^c$ can be read off a complete NLO calculation
for a specific process.

Expanding the resummed cross section through NNLO and matching
with the NLO result gives us the master formula for the NNLO corrections 
\cite{NKuni}
\beqa
{\hat{\sigma}}^{(2)}&=&
\sigma^B \frac{\alpha_s^2(\mu_R^2)}{\pi^2} \;
\frac{1}{2} \, c_3^2 \; {\cal D}_3(s_4)
\nonumber \\ && \hspace{-18mm}
{}+\sigma^B \frac{\alpha_s^2(\mu_R^2)}{\pi^2} \;
\left\{\frac{3}{2} \, c_3 \, c_2 - \frac{\beta_0}{4} \, c_3
+\sum_j C_j \, \frac{\beta_0}{8}\right\} \; {\cal D}_2(s_4)
+\frac{\alpha_s^{d_{\alpha_s}+2}(\mu_R^2)}{\pi^2} \;
\frac{3}{2} \, c_3 \, A^c\; {\cal D}_2(s_4)
\nonumber \\ && \hspace{-18mm}
{}+\sigma^B \frac{\alpha_s^2(\mu_R^2)}{\pi^2}  
\left\{ c_3 c_1 +c_2^2
-\zeta_2  c_3^2 -\frac{\beta_0}{2}  T_2
+\frac{\beta_0}{4}  c_3  \ln\left(\frac{\mu_R^2}{M^2}\right)
+c_3 \frac{K}{2}
-\sum_j\frac{\beta_0}{4}  {B'}_j^{(1)} \right\} 
{\cal D}_1(s_4)
\nonumber \\ && \hspace{-18mm}
{}+\frac{\alpha_s^{d_{\alpha_s}+2}(\mu_R^2)}{\pi^2} \;
\left\{\left(2\, c_2-\frac{\beta_0}{2}\right)\, A^c+c_3 \, T_1^c
+F^c\right\} \; {\cal D}_1(s_4)
+{\cal O}\left({\cal D}_0(s_4)\right)\, ,
\eeqa
where we show terms explicitly through NNLL.
Here
$F^c={\rm tr} [H^{(0)} \left({\Gamma'}_S^{(1)\,\dagger}\right)^2 S^{(0)}
+H^{(0)} S^{(0)} ({\Gamma'}_S^{(1)})^2
+2 H^{(0)} {\Gamma'}_S^{(1)\,\dagger} S^{(0)} {\Gamma'}_S^{(1)}]$.

Finally, expanding the resummed formula through NNNLO  
and matching with the NLO and NNLO results gives us the
master formula for the NNNLO corrections
\beqa
{\hat{\sigma}}^{(3)}&=&
\sigma^B \frac{\alpha_s^3(\mu_R^2)}{\pi^3} \;
\frac{1}{8} \, c_3^3 \; {\cal D}_5(s_4)
\nonumber \\ && \hspace{-23mm}
{}+\sigma^B \frac{\alpha_s^3(\mu_R^2)}{\pi^3} \;
\left\{\frac{5}{8} \, c_3^2 \, c_2 -\frac{5}{2} \, c_3 \, X_3\right\} \;
{\cal D}_4(s_4)
+\frac{\alpha_s^{d_{\alpha_s}+3}(\mu_R^2)}{\pi^3} \;
\frac{5}{8} \, c_3^2 \, A^c \; {\cal D}_4(s_4)
\nonumber \\ && \hspace{-23mm}
{}+\sigma^B \frac{\alpha_s^3(\mu_R^2)}{\pi^3}
\left\{c_3 \, c_2^2 +\frac{1}{2} \, c_3^2\, c_1
-\zeta_2 \, c_3^3 +(\beta_0-4c_2) \, X_3 +2 c_3 \, X_2
-\sum_j C_j \, \frac{\beta_0^2}{48}\right\} {\cal D}_3(s_4)
\nonumber \\ && \hspace{-23mm}
{}+\frac{\alpha_s^{d_{\alpha_s}+3}(\mu_R^2)}{\pi^3} 
\left\{\frac{1}{2} c_3^2 T_1^c+\left[2 c_3  c_2
-\frac{\beta_0}{2}  c_3 -4\,  X_3 \right] A^c +c_3  F^c\right\} 
{\cal D}_3(s_4)+{\cal O}\left({\cal D}_2(s_4)\right)
\eeqa
where again we show terms explicitly through NNLL.
Here
$X_3=(\beta_0/12) c_3-\sum_j C_j \beta_0/24$ and 
$X_2=-(\beta_0/4)T_2+(\beta_0/8)c_3 \ln(\mu_R^2/M^2)
+c_3K/4-\sum_j(\beta_0/8){B'}_j^{(1)}$.

This calculation has recently been applied to charged Higgs production
with a top quark via bottom gluon fusion at the LHC through NNLO \cite{NKch}, 
where for a charged Higgs mass of 500 GeV the NLO-NLL soft-gluon corrections 
provide a enhancement of 38\% over the leading-order cross section 
and the NNLO-NLL corrections provide a enhancement 
of 11\% over the NLO-NLL cross section. 
A new calculation of the NNNLO-NLL corrections shows that they 
provide an additional 7\% 
enhancement over the NNLO-NLL cross section and further stabilize the
scale dependence of the cross section.
Similarly, for top quark pair production at the Tevatron the scale 
dependence is considerably decreased. More details will be given 
in a forthcoming paper.

\end{document}